\documentclass[final,5p,times,twocolumn]{elsarticle}

\usepackage{amssymb}
\usepackage{breqn}
\usepackage{xcolor}
\usepackage{mathtools}

\begin{document}
	
	\begin{frontmatter}
		

		\title{Noise induced extreme events in single Fitzhugh-Nagumo oscillator}
		
		\author[inst1]{S. Hariharan}
		\author[inst1]{R. Suresh}
		\author[inst1]{V. K. Chandrasekar}
		\affiliation[inst1]{organization={Department of Physics},
			addressline={Centre for Nonlinear Science and Engineering, School of Electrical and Electronics Engineering,\\ SASTRA Deemed University}, 
			city={Thanjavur},
			postcode={613402}, 
			state={Tamilnadu},
			country={India}}
		
		\begin{abstract}
			The FitzHugh-Nagumo (FHN) model serves as a fundamental neuronal model which is extensively studied across various dynamical scenarios, we explore the dynamics of a scalar FHN oscillator under the influence of white noise. Unlike previous studies, in which extreme events (EE) were observed solely in coupled FHN oscillators, we demonstrate that a single system can exhibit EE induced by noise. Perturbation of the deterministic model in its steady state by random fluctuations reveals the emergence of subthreshold/small-amplitude oscillations (SAO), eventually leading to rare and extreme large-amplitude oscillations (LAO), which become particularly evident at minimal noise intensities. We elucidate the route by which these EE emerge, confirming their occurrence through probability calculations of trajectories in phase space. Additionally, our investigation reveals bursting phenomena in the system, which are characterized by specific levels of noise amplitude and elucidated using inter-spike interval statistics. At higher noise amplitudes, frequent LAO production is observed and attributed to self-induced stochastic resonance. The emergence of EE is explained through the theory of large fluctuations, with the escape rates of trajectories estimated via both analytical and numerical approaches. This study is significant because it reveals EE and bursting phenomena in a single FHN oscillator, offering potential new insights into the dynamics of neuronal populations.
		\end{abstract}
		%
		
		
	\end{frontmatter}
	
	
	\section{Introduction}
	\label{sec:sample1}
	
	
	The study of noise-induced dynamics in nonlinear systems is a crucial area of research, as nearly all real-world systems, ranging from biological systems to mechanical, electrical, and financial systems, are subject to random perturbations. These noisy forces can significantly alter a system’s behavior, often resulting in unexpected phenomena such as noise-induced order \cite{matsumoto1983noise,Gassman1997noise}, chaos \cite{werner1984noise}, and transitions between stable states \cite{gao1999noise_chaos}. Various systems, including communication networks, power grids, environmental processes, financial markets, and neural networks, are heavily influenced by stochastic forces, making it essential to understand the interplay between noise and system dynamics \cite{duan2015introduction, tsimring2014noise,fishmm1990electrical}.
	
	Noise in neuronal systems arises from multiple sources, such as fluctuations in ion channel conductance, synaptic activity, and environmental disturbances. These noise sources significantly influence the dynamics of neuronal oscillators, influencing spike timing, synchronization, and signal transmission within neural networks \cite{faisal2008noise}. Noise-induced variability is critical to understanding normal and pathological neural functions, such as epilepsy, in which abnormal neuronal synchronization can lead to extreme events (EE) like seizures. \cite{osorio2010,lehnertz2006}. Moreover, studying stochastic forces is crucial for developing neuroprosthetics and brain-computer interfaces, as it improves the interpretation of inherently noisy neural signals \cite{lebedev2006}. In artificial neural networks (ANNs), stochastic techniques such as dropout and stochastic gradient descent mimic the beneficial variability found in biological systems, enhancing training efficiency and model generalization \cite{bottou2010}. A deeper understanding of noise-driven dynamics is, therefore, essential for advancing both neuroscience and machine learning.
	
The pioneering work of Hodgkin and Huxley \cite{hodgkin1952} laid the foundation for mathematical models that capture neuronal dynamics, which were later simplified in models like the Hindmarsh-Rose (HR), FitzHugh-Nagumo (FHN), and Morris-Lecar models. The FHN model \cite{fitzhugh1961}, a reduction of the Hodgkin-Huxley model, captures the core mechanisms of excitable systems, including spike generation and recovery, and has been applied to cellular signaling \cite{sgro2015intracellular}. A comprehensive review of the FHN model’s applications over six decades is presented in \cite{cebrian2024six}. Due to its simplicity, the FHN model has become a key tool for investigating noise-induced phenomena in neurons, such as noise-induced spiking \cite{makarov2001spiking}, coherence resonance (CR) \cite{pikovsky1997coherence}, and stochastic resonance (SR) \cite{longtin1997sr}. Recent studies have also explored more complex behaviors in the FHN model, including mixed-mode oscillations (MMO) \cite{muratov2008mmo}, noise-induced stabilization \cite{manchein2022noise}, and transitions to chaotic dynamics \cite{bashkirtseva2014noise}. Self-induced stochastic resonance (SISR) has been identified as a unique, noise-driven resonance phenomenon \cite{muratov2005sisr, deville2005sisr}. Recent research has extended resonance dynamics to multilayer FHN networks \cite{wu2024resonance}, and a generalized FHN model with tristable dynamics has been studied under white Gaussian noise \cite{xu2020dynamics}. Despite these advancements, a critical gap remains in the study of EE in single-neuron models like the FHN, particularly regarding noise-induced transitions to EE. This gap represents an important area for further investigation.
	
Extreme events are rare, high-amplitude deviations from a system’s typical behavior, observed across diverse domains, including social dynamics \cite{helbing2001social}, oceanography \cite{chabchoub2011ocean}, optics \cite{bonatto2011optics}, geophysics \cite{sornette2006geo}, economics \cite{bunde2012eco}, and power grid failures \cite{dobson2007power}. In neuroscience, EE are associated with pathological states like epileptic seizures, characterized by excessive synchronized neuronal firing \cite{osorio2010, lehnertz2006}. Studying EE in a single FHN neuron model is particularly significant, as it offers a controlled setting to explore the role of noise in the emergence of EE. Single-neuron models enable detailed bifurcation and fluctuation analyses, clarifying the mechanisms underlying EE while avoiding complexities introduced by network interactions. This approach also provides a baseline for understanding larger-scale network dynamics, as individual neuronal behaviors critically influence collective phenomena like synchronized bursts. Noise-induced effects, such as stochastic resonance and bursting, highlight the importance of studying isolated neurons, with potential applications in predicting pathological activity like seizures \cite{osorio2010}. Moreover, understanding EE in single neurons sheds light on foundational neuronal processes like bursting and epileptic discharges, which are critical for neuronal information processing and the broader dynamics of neural networks \cite{Gupta2020SingleNeuron}.

While much research has explored EE in coupled neural systems, relatively few studies have examined how noise alone induces EE in individual neurons \cite{hariharan2024noise}. For instance, factors like multistability and super-EE have been observed in coupled Hindmarsh-Rose models \cite{vijay2023ses}, and electrical coupling has been shown to trigger EE with complex statistical properties \cite{olenin2023hree}. In coupled FHN models, mechanisms such as time-delay and network topology drive EE \cite{karnatak2014fnee, saha2017fndelay}. However, the potential for noise-induced EE in single, uncoupled neurons remains underexplored. Addressing this gap, our study investigates the FHN model with noise as the primary mechanism for inducing EE, contributing novel insights into how stochastic dynamics drive such rare events.
	
	A primary contribution of our study is demonstrating that noise, without coupling or external forcing, can induce EE in a single FHN oscillator. By selecting parameters that position the system in a non-oscillating regime--leading to a steady state in the absence of noise--we analyze how stochastic perturbations alter the system’s behavior. Our findings indicate that noise alone can drive the system into an EE regime, resulting in rare, large excursions in phase space. This challenges the common view that EE typically arise in systems with deterministic instability or coupling and highlights new possibilities for studying EE in isolated, noise-driven systems.
	
	We conduct a bifurcation analysis to unravel the mechanisms behind noise-induced EE, revealing how stochastic perturbations reshape the FHN model’s dynamics.  We find that sub-threshold oscillations (SAO) can be amplified by noise, resulting in large-amplitude oscillations (LAO) that qualify as EE. This transition is confirmed by calculating the probability distribution of trajectory excursions in phase space, illustrating the profound impact of noise on the system’s limit cycle dynamics.
	
	Our study also investigates self-induced stochastic resonance (SISR) , where noise, without periodic forcing, induces coherent oscillations in the system which is studied in two-dimensional models. We show that noise-induced extreme events (EE) in the FHN model are intricately dependent on noise intensity, which acts as a key control parameter in the emergence of oscillations. Additionally, by applying the theory of large fluctuations, we calculate the escape rates of system trajectories both analytically and numerically, enhancing our understanding of the stochastic dynamics underlying EE. Our analyses reveal that small noise-induced perturbations can cause the system’s trajectories to escape from a stable steady state, leading to significant excursions in phase space. By clarifying the role of noise in inducing EE in neuronal oscillators, our work may provide valuable insights for predicting and managing extreme neuronal activity in conditions such as epilepsy and migraines. Additionally, our work aims to bridge the gap between the dynamics of individual neurons and the collective behavior of neural networks. We systematically uncover the influence of noise and nonlinearities through a detailed examination of a single neuron. These findings serve as a robust foundation for extending the research to more complex and biologically realistic network models.
	
	In summary, this study investigates the generation of EE in a single FHN oscillator driven by noise. By analyzing the system’s response to stochastic perturbations, we reveal how noise induces large-amplitude excursions in phase space, leading to EE. Our findings highlight the importance of noise in modulating neuronal dynamics and provide new insights into the mechanisms driving extreme behaviors in excitable systems. The structure of this paper is as follows: Sec. 2 presents the FHN model and the noise-induced dynamics framework. Sec. 3 and 4 discuss extreme events and bifurcation analysis. Sec. 5 examines noise-induced bursting, while Sec. 6 explores resonance phenomena. Finally, Sec. 7 summarizes the key findings and their implications.
	
	\section{System description and the emergence of extreme events}
	In this study, we focus on a single unit of the FHN model perturbed by white Gaussian noise in the membrane potential $x$. The system dynamics are governed by the following equations:
	
	\begin{eqnarray}
		\label{eq:e2}
		\dot{x} &=& x(a-x)(x-1)-y+\sqrt{2D}\xi(t),\\ 
		\dot{y} &=& bx-cy \nonumber,
	\end{eqnarray} 
	\begin{figure*}[h!]
		\centering
		\includegraphics[width=1.0\linewidth]{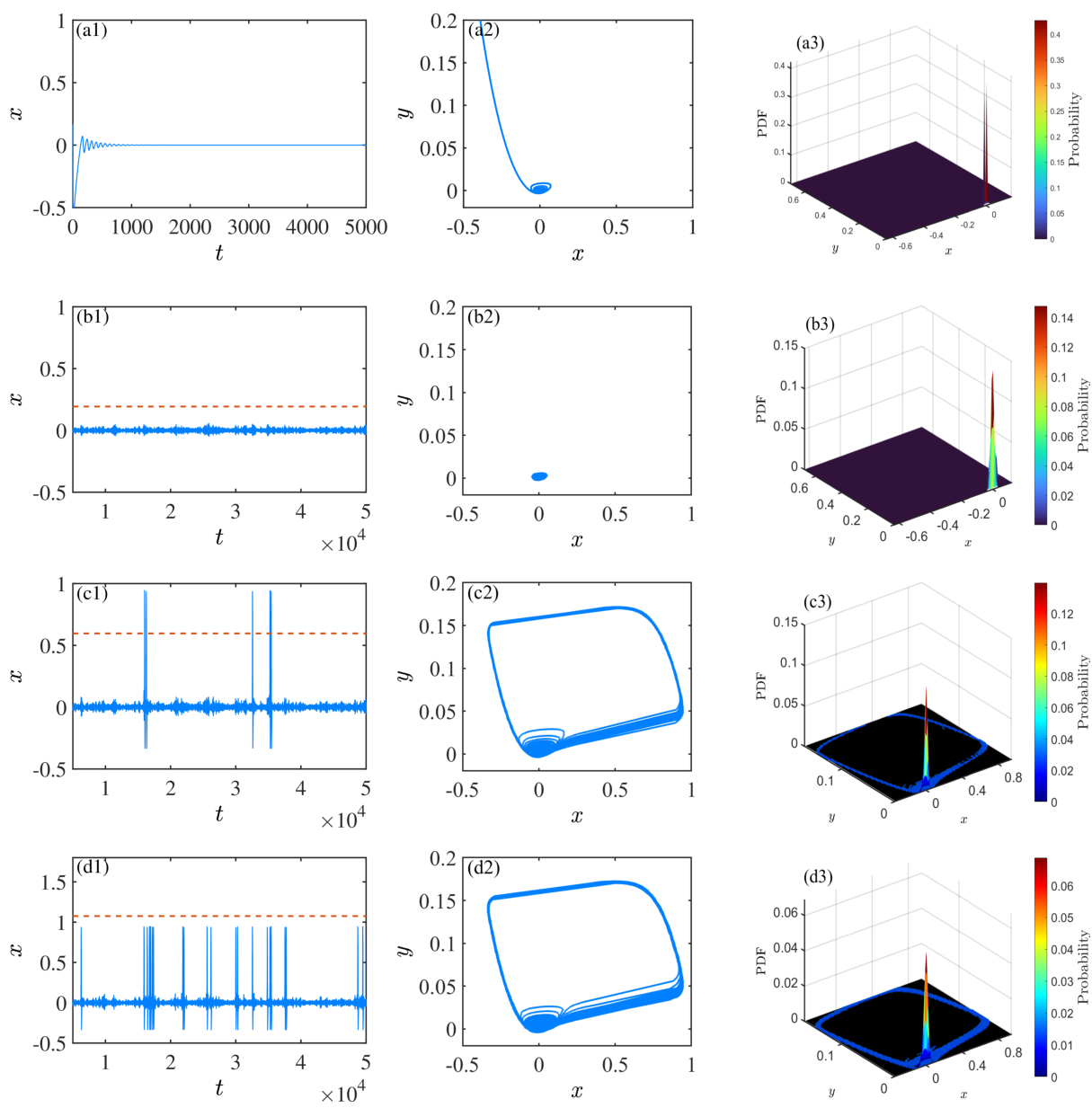}
		\caption{\label{feu_st} The left panel shows the time series, the middle panel presents the corresponding phase portrait, and the right panel displays the probability distribution function (PDF) of the system variables as a function of noise intensity $D$. The dashed line indicates the threshold $H_s$, calculated from the time series. The system parameters are fixed at $a = -0.012$, $b = 0.00702$, and $c = 0.02$. Figures (a1-a3) illustrate the deterministic case ($D = 0$), where the system evolves towards a steady state. In contrast, when the system is subjected to a small noise intensity ($D = 0.0001$), it exhibits small amplitude oscillations (SAO) as shown in (b1-b3). Panels (c1-c3) depict the occurrence of rare, recurrent large amplitude oscillations (LAO), where oscillations exceeding the dashed line are identified as extreme events (EE) for $D = 0.0002$. Finally, (d1-d3) demonstrates the frequent appearance of LAO alongside SAO for $D = 0.0003$. Note that in this case, none of the LAOs exceed the dashed line, and therefore, they are not classified as EEs.}
	\end{figure*}
	where $x$ and $y$ denote the membrane potential and recovery variable, respectively. $D$ represents the noise intensity of the Gaussian white noise $\xi(t)$, and $a, b$, and $c$ are system parameters.
	
	In the absence of noise ($D = 0$), the system possesses equilibrium points: 

		\begin{equation}
			\begin{split}
				(x_1, y_1) &= (0, 0), \\ 
				(x_{2,3}, y_{2,3}) & = \left( \frac{1}{2}\left(1 + a \pm \frac{\sqrt{-4b + c - 2ac + a^2c}}{\sqrt{c}}\right), \right. \\
				&\quad \left. \frac{b}{2c}\left(1 + a \pm \frac{\sqrt{-4b + c - 2ac + a^2c}}{\sqrt{c}}\right) \right).\nonumber
			\end{split}
		\end{equation}

	with their stability depending on the system parameters.
	
	Since our focus is on noise-induced excitability, we choose parameter values from the non-self-oscillating regime: $a = -0.012$, $b = 0.007$, and $c = 0.02$. For these values, the system converges asymptotically to a stable steady state. The time series and phase portrait for this deterministic case, shown in Figs. \ref{feu_st}(a1) and \ref{feu_st}(a2), depict a stable spiral trajectory converging toward the fixed point. Additionally, the probability density function (PDF) of the trajectory in the $(x, y)$ plane is illustrated in Fig. \ref{feu_st}(a3), with the color bar indicating probability density, which peaks near the stable equilibrium. When noise is introduced, the system begins to oscillate around the equilibrium. For small noise intensities (e.g., $D = 0.0001$), the system exhibits small amplitude oscillations (SAO), also referred to as sub-threshold oscillations. Figures \ref{feu_st}(b1-b3) show the corresponding time series, phase portrait, and PDF. The phase portrait reveals small excursions in the phase space around the equilibrium, and the PDF broadens slightly compared to the deterministic case.
	
	As the noise intensity increases, the amplitude of SAO grows. At certain noise levels, the system occasionally displays large amplitude oscillations (LAO) amidst the SAO, occurring at random intervals. These rare, significantly large events are classified as EE. For example, at $D = 0.0002$, Figs. \ref{feu_st}(c1-c2) illustrate the time evolution and phase portrait of the system, respectively, showing the occurrence of LAO alongside SAO. The PDF demonstrates the probability of both types of oscillations in the phase space. As noise intensity increases further, the frequency of LAO increases, as shown in Figs. \ref{feu_st}(d1-d3) for $D = 0.0003$, where LAO appears more frequently. With even higher values of $D$, the system displays increasingly frequent large amplitude oscillations.
	
	To classify EE, we estimate a predefined threshold height, $H_T = \langle P_n \rangle + N\sigma$, where $\langle P_n \rangle$ represents the mean of the maxima, $N$ is a scaling factor (typically $N \in [2, 8]$ depending on the system), and $\sigma$ is the standard deviation \cite{chowdhury2022review}. For this study, we set $N = 8$. Given the system’s stochastic nature, identifying true maxima and minima is not trivial. To address this, we apply an algorithm for detecting local maxima \cite{MATLABfindpeaks}, specifying a minimum peak amplitude $P_T = 0.001$ and a minimum distance of $t = 70$ between consecutive peaks, which corresponds to the minimum pulse width in the system. This approach effectively filters out minor noise-induced peaks while capturing significant maxima. The threshold $H_T$ is shown as a dashed line in Fig. \ref{feu_st}, with peaks that exceed this threshold classified as EE. For a noise intensity of $D = 0.0001$ (Fig. \ref{feu_st}(b1)), no oscillations exceed the threshold $H_T$, indicating the absence of EE. At $D = 0.0002$ (Fig. \ref{feu_st}(c1)), several peaks cross the threshold, signaling the occurrence of EE. However, at $D = 0.0003$ (Fig. \ref{feu_st}(d1)), although large-amplitude oscillations (LAO) become more frequent, none exceed the threshold, and thus no EE are observed at this noise level.
	
	\begin{figure}[h!]
		\centering
		\includegraphics[width=1.1\linewidth]{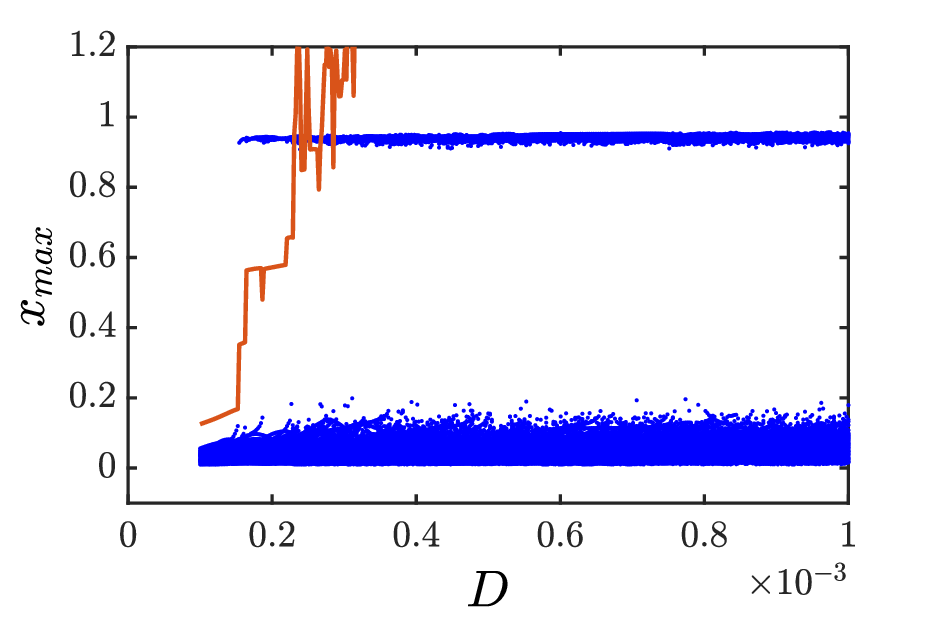}
		\caption{\label{d_bif} Bifurcation diagram of noise intensity ($D$), illustrating the system’s dynamic behavior. The red curve represents the threshold height ($H_T$) used to classify events as either extreme or normal. }
	\end{figure}
	
	To further analyze the system’s behavior, we present the bifurcation diagram of the variable  $x_{\text{max}}$  as a function of noise intensity ($D$) in Fig. 2.  In this diagram, the blue 
		(dark gray) dots represent the maxima of the  x-variable, while the red line denotes the threshold height  $H_T$. For  $D = 0.0001$, the system  oscillates near the equilibrium point without producing spikes, consistent with the time series shown in Fig. 1(b1). This SAO behavior persists for  $D \in (0.0001, 0.00014]$.  At  $D = 0.00015$, the system transitions to rare spiking 
		events, with these spikes exceeding the threshold height $H_T$.  These spikes are classified as EE and can observed in the time series of Fig. 1(c1) for $D = 0.0002$. The EE behavior continues for 
		$D \in (0.00015, 0.00023]$.
	
 		As $D$ increases further, the infrequent nature of spiking is lost, and the spikes become more frequent over time. In other words, the system transitions from EE state to a frequent 
		bursting state, with intermittent occurrences of EE during this transition. This intermediate state is evident in Fig. 2, where the $H_T$ line exhibits more fluctuations for $D$ values between 0.00023 and 0.00029. For $D >= 0.0003$, the system no longer exhibits peaks exceeding $H_T$, indicating the cessation of EE.  The dynamics remain consistent with no EE for $D>0.0003$, as depicted in the times series of Fig. 1(d1).
	
	\begin{figure}[h!]
		\centering
		\includegraphics[width=1.1\linewidth]{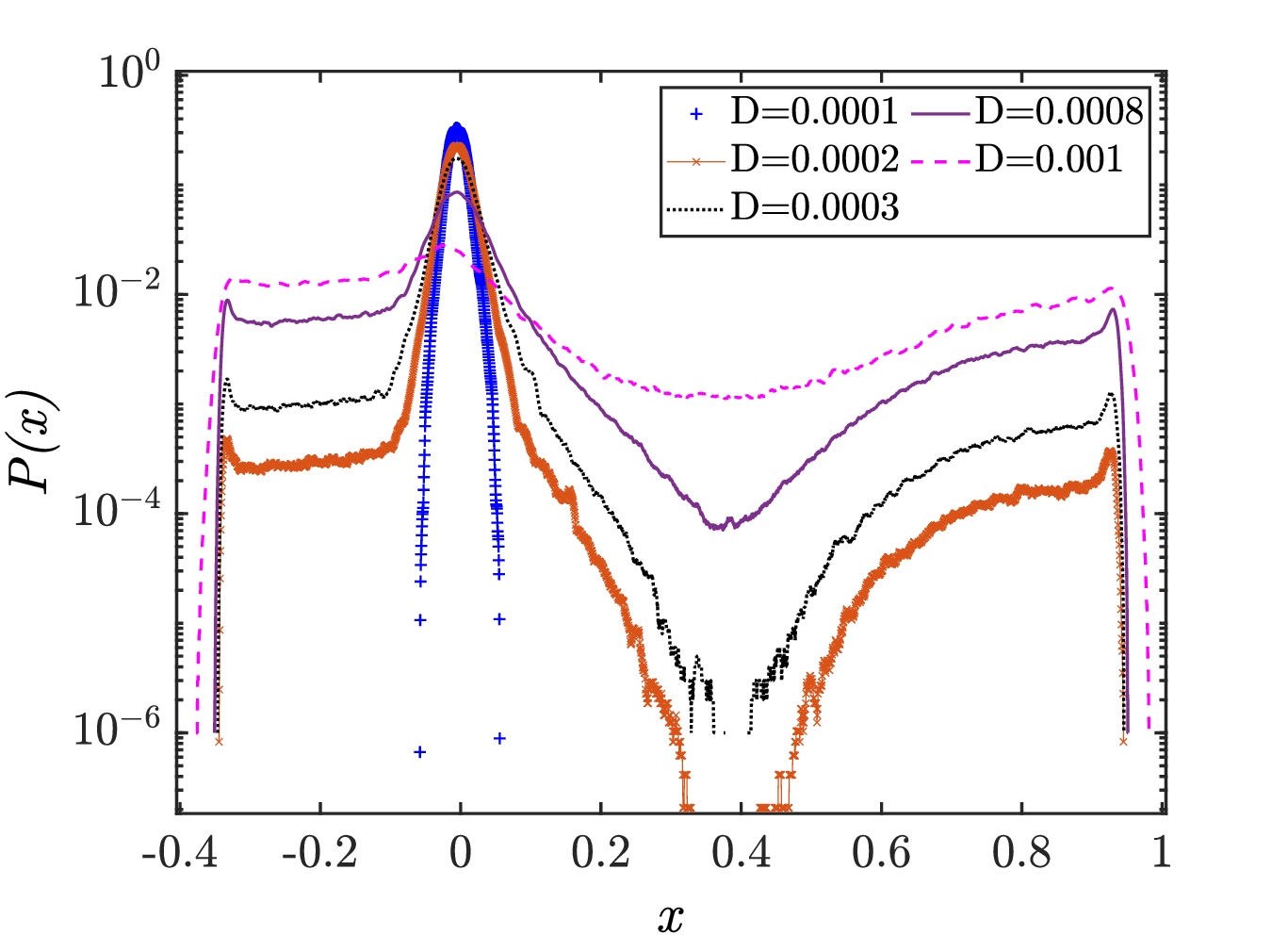}
		\caption{\label{p(x)}The probability distribution of the system variable $x$ for various noise intensities $D$ shows that the system undergoes P-bifurcation.}
	\end{figure}

	The defining characteristic of EE is their rarity, combined with their substantial impact on the system. To confirm the infrequent occurrence of these events, whether low or high amplitude oscillations, various statistical methods are used for analysis. In this study, we estimate the probability distribution of the state variable $x$, denoted as $P(x)$, under different noise intensities ($D$). The resulting $P(x)$ distributions, presented in Fig. \ref{p(x)} as semi-log plots for various $D$ values, reveal intriguing patterns.
	
	For $D = 0.0001$, the $x$-variable oscillates around the equilibrium point, producing SAO. The corresponding probability distribution, $P(x)$ (denoted by the “+” symbols), is unimodal and concentrated near $x = 0$, indicating that the system predominantly remains close to the equilibrium. However, when the noise intensity increases to $D = 0.0002$, although the system still spends most of its time near the equilibrium point, the probability distribution (denoted by the “×” symbols) shifts. A new peak emerges in the sub-threshold region, but its probability is lower than that observed at $D = 0.0001$ due to the appearance of another mode near $x \approx 0.9$ in the supra-threshold region. This new mode is associated with the rare occurrence of LAO. This transition from an unimodal to a bimodal distribution as the noise intensity increases represents a significant change in the system’s behavior, known as a P-bifurcation \cite{arnold1995rds}, which occurs in noise-induced systems when EE are present. As the noise intensity increases further, for example, at $D = 0.0003$, $0.0008$, and $0.001$, the probability distribution widens, with $P(x)$ in the subthreshold region decreasing and increasing in the supra-threshold region. These broader distributions are illustrated in Fig. \ref{p(x)} as dotted, continuous, and dashed lines, respectively. It is important to note that the system undergoes P-bifurcation only during the transition from SAO to noise-induced EE at $D = 0.00015$ as indicated in Fig. \ref{d_bif}. After the bifurcation, the occurrence of LAO becomes more frequent as $D$ increases further \cite{bashkirtseva2018methods,hariharan2024noise}.
	
	\section{Bifurcation analysis for the system parameters}
\begin{figure*}[h!]
		\centering
		\includegraphics[width=1.02\linewidth]{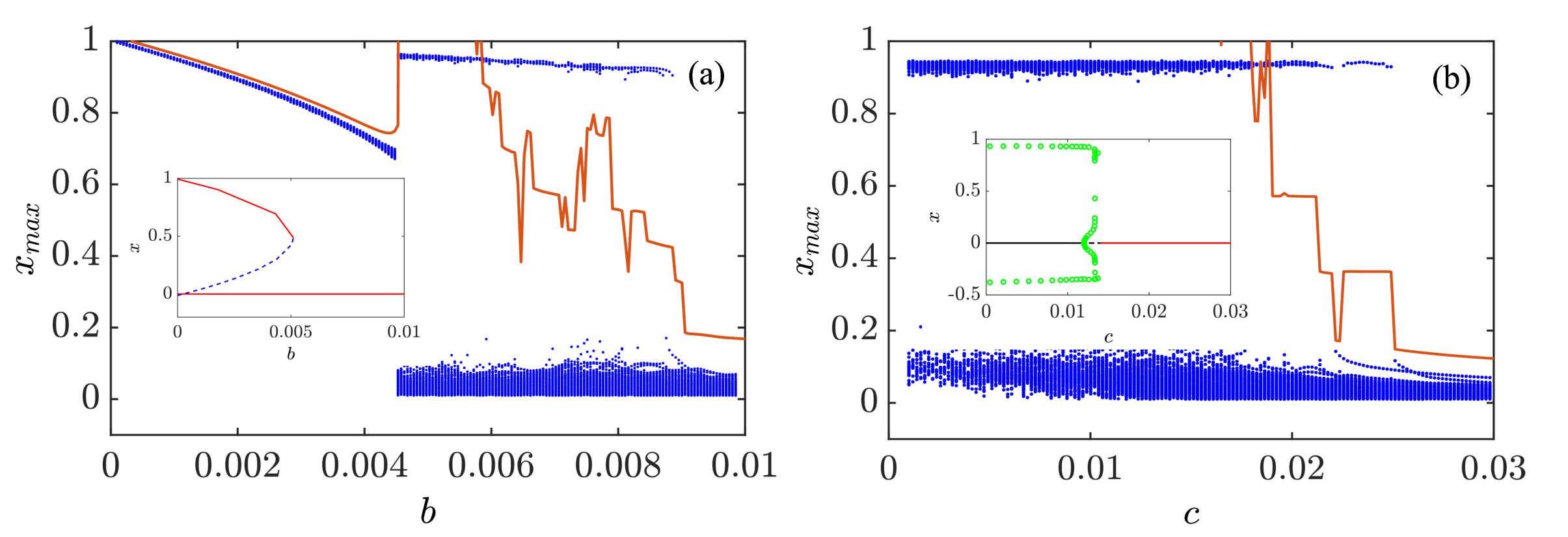}
		\caption{\label{bifur}The bifurcation diagram illustrates the system's behavior as a function of two parameters: (a) $b$, varying within the range of (0.0, 0.01), and (b) $c$, within the range of (0.0, 0.03), highlighting the emergence of extreme events. The red curve in both (a) and (b) denotes the $H_T$ value. The noise intensity is held constant at $D = 0.0002$. Insets depict the stability of fixed points under noise-free conditions.  }
	\end{figure*}

	To investigate the mechanism of the emergence of EE, we perturb the system with weak noise ($D = 0.0002$) and conduct a bifurcation analysis of the system parameters $b$ and $c$. Figure \ref{bifur}(a) shows the bifurcation diagram for $b$ with noise, where the red continuous line represents the threshold height, $H_T$. At $b = 0$, the system initially undergoes a transient phase before fluctuating near $x = 1$. As $b$ increases, these fluctuations progressively shift to lower $x$ values. To better understand this behavior, we also examine the bifurcation of $b$ in the deterministic case ($D = 0$), depicted in the inset of Fig. \ref{bifur}(a).
	
	In the deterministic system, the red line indicates a stable steady state, while the dashed blue line represents an unstable state. At $b = 0$, the system shows multistability: if the initial conditions (ICs) are near the origin, the system stabilizes at the fixed point $(0,0)$. However, if the ICs are set farther from the origin (e.g., $x = 0.8$), the system stabilizes near $x = 1$. This multistability continues until $b = 0.0048$, where the system undergoes a saddle-node bifurcation as the saddle and node collide. For $b > 0.0048$, the system consistently reaches the steady state at $(x, y) = (0, 0)$, regardless of the ICs.
	
	When noise is introduced, this deterministic behavior changes. For $b = 0$, the system now fluctuates near $x = 1$, but these fluctuations are independent of the ICs due to the noise perturbation. This state is termed a ``fluctuating steady state” as no significant oscillations occur. This behavior persists until $b = 0.0048$, at which point the noise-induced bifurcation mirrors the deterministic one. However, for $b > 0.0048$, the system’s dynamics diverge from the deterministic case. Instead, the system begins to produce oscillations with varying amplitudes, driven by the interplay between noise and the parameter $b$. Within the range $b \in (0.0048, 0.0064)$, the system exhibits frequent LAO that do not meet the threshold condition for EE. As $b$ increases to $b = 0.0065$, fewer LAOs occur, but they now satisfy the threshold, leading to the generation of EE. This EE regime persists for $b \in (0.0065, 0.009)$. Beyond this range, the system returns to random small oscillations, converging towards the steady state predicted by the deterministic bifurcation. Due to the noise, SAO occurs around the fixed point for $b \in (0.009, 0.010)$.
	
	Figure \ref{bifur}(b) presents the bifurcation diagram for the parameter $c$, with the inset illustrating the bifurcation in the deterministic case ($D = 0$). In the deterministic scenario, as $c$ decreases (moving from right to left in the inset figure), the system’s fixed point remains stable within the range $c \in (0.014, 0.03)$, as shown by the red line, leading to a steady state regardless of the ICs. However, when $c$ falls below $0.014$, the fixed point becomes unstable, indicated by the dashed black line. At this point, the system’s behavior becomes dependent on the ICs: if the ICs are close to the equilibrium point $(x_0, y_0) = (0, 0)$, the system will still converge to a steady state. Otherwise, the system transitions to an oscillatory state. This shift in dynamics is driven by a super-critical Hopf bifurcation, where the stable equilibrium loses stability, marking the onset of oscillatory behavior. This bifurcation also reveals multistability within the range  $c \in (0.013, 0.0)$.
	
	Under the influence of noise, the system’s behavior significantly changes. For $c \in (0.030, 0.0251)$, the system exhibits non-uniform SAO around the equilibrium point. As $c$ decreases to $0.025$, the system departs from the vicinity of the fixed point, and its trajectories explore a larger region of the phase space, resulting in LAO. This expansion of the attractor above the threshold height $H_T$ satisfies the condition for EE generation. In this regime, SAO and LAO alternate unpredictably, with noise inducing an early bifurcation that mirrors the deterministic system’s behavior. As $c$ continues to decrease, the system becomes highly sensitive, producing numerous LAO in the range $c \in (0.0249, 0.018)$. Eventually, for $c \in (0.0179, 0.0)$, the system displays alternating small and frequent large oscillations, reflecting the overall dynamics shaped by both noise and the system parameters.
	
	\section{Noise-induced bursting}
At low noise amplitudes (D = 0.0003), the system begins to generate spikes that deviate from the characteristics typically associated with EE. As the noise intensity increases beyond D = 0.0004, a new dynamical phenomenon, known as noise-induced bursting, emerges. This behavior has been well-documented in prior studies \cite{channell2009burst, ryashko2020burst, lopez2023controlling}. Additionally, similar bursting phenomena have been observed in the FHN model under the influence of quasi-monochromatic noise \cite{baltanas1998burst}. Notably, when white Gaussian noise is modified into Ornstein-Uhlenbeck noise ($\eta(t)$), it induces bursting that manifests as the skipping of spikes, as described in \cite{longtin1995synchronization}. These findings underscore the intricate interplay between noise and the system’s inherent dynamics in producing complex neuronal behaviors.
	\begin{figure*}[h!]
		\centering
		\includegraphics[width=0.75\linewidth]{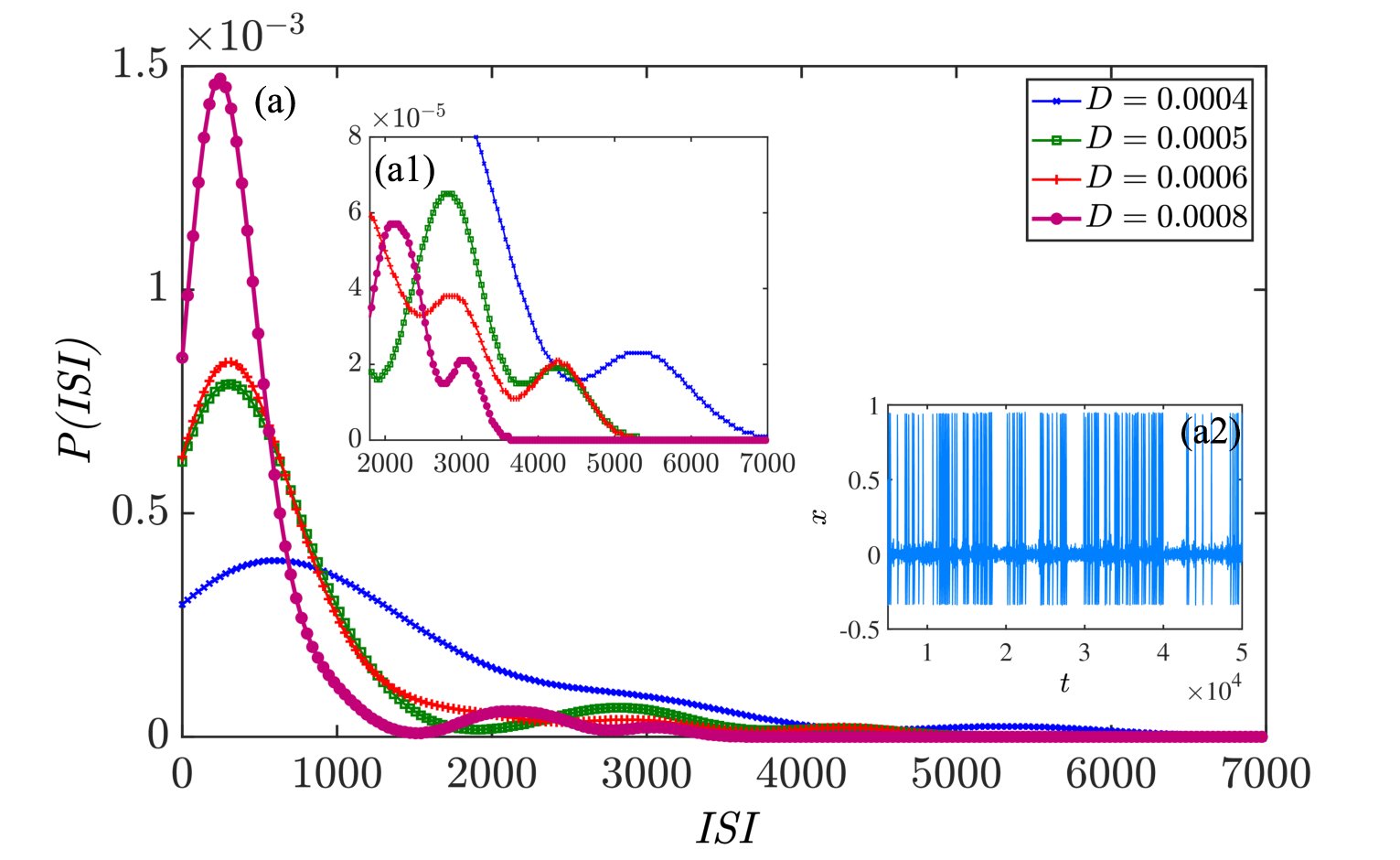}
		\caption{\label{isi_b}Probability distribution of inter-spike intervals (ISI) illustrating the behavior of bursting oscillations across varying noise intensities. Inset (a1) provides a magnified view of the ISI distribution, highlighting changes within the 2000 to 7000 range. Inset (a2) displays the dynamics of noise-induced bursting oscillations for a noise intensity of  D = 0.0008. }
	\end{figure*}
	\par
	
	To quantify the system’s dynamics, we analyzed the probability distribution of inter-spike intervals (ISI) at different noise intensities, $D$, as shown in Fig. \ref{isi_b}. Spikes were identified when the system’s state variable exceeded the threshold of $x > 0.8$. At $D = 0.0004$ (blue $*$ curve), the ISI distribution reveals a broad peak at shorter intervals, although the overall probability remains low. A secondary peak is observed at longer intervals (inset, Fig. \ref{isi_b}(a1)), reflecting the presence of SAO between spikes. LAO tends to occur after longer time intervals.
	
	At $D = 0.0005$ (green $\blacksquare$ curve), the ISI distribution narrows in the short-interval range, with an increased overall probability. A secondary peak in the 2000–4000 range indicates quiescent periods dominated by SAO between bursts. When $D$ is increased to 0.0006 (red $+$ curve), the ISI distribution remains similar, with high probabilities for short intervals and minimal change in dynamics compared to $D = 0.0005$.
	
	As $D$ increases to 0.0008 (magenta $\medbullet$ curve), the ISI distribution exhibits significant shifts. The distribution becomes more concentrated around shorter intervals, and two low-probability peaks emerge in the 1500–5000 range. The first peak, with the highest probability, corresponds to short ISI bursts, while the lower-probability peaks reflect noisy quiescence periods. This pattern indicates irregular, frequent bursting events, as shown in the inset of Fig. \ref{isi_b}(a2) for $D = 0.0008$. The system exhibits alternating small and frequent large-amplitude oscillations that are aperiodic and irregular, with significantly reduced quiescence times due to increased noise intensity. This behavior is characteristic of noise-driven stochastic bursting oscillations.
	
	\section{Resonance dynamics}
	\begin{figure}[h!]
		\centering
		\includegraphics[width=1.0\linewidth]{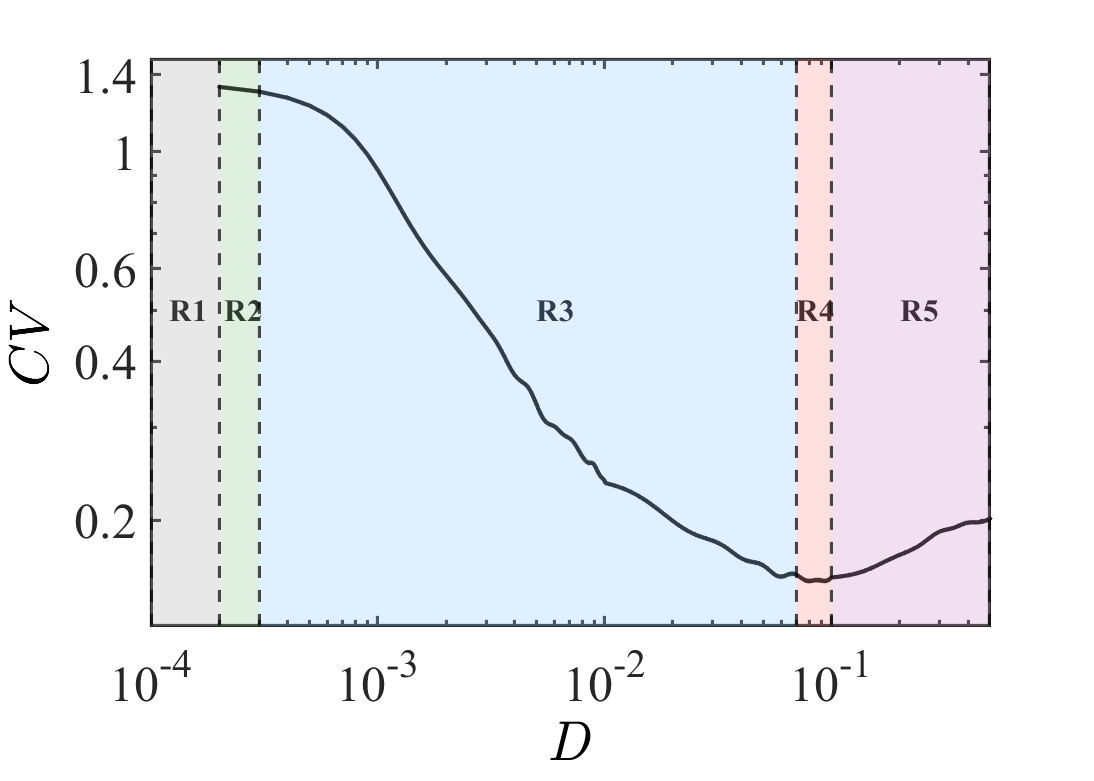}
		\caption{\label{cv}Coefficient of variation (CV) as a function of noise intensity $D$, shown on a logarithmic scale, highlights distinct behavioral regions. In region R1, no spikes are detected. Region R2 corresponds to the occurrence of EE, followed by consistent, random, and infrequent spiking in region R3. Region R4 illustrates the emergence of SISR behavior, while region R5 displays incoherent spiking as noise intensity ($D$) increases further.}
	\end{figure}
	\begin{figure}[h!]
		\centering
		\includegraphics[width=1.0\linewidth]{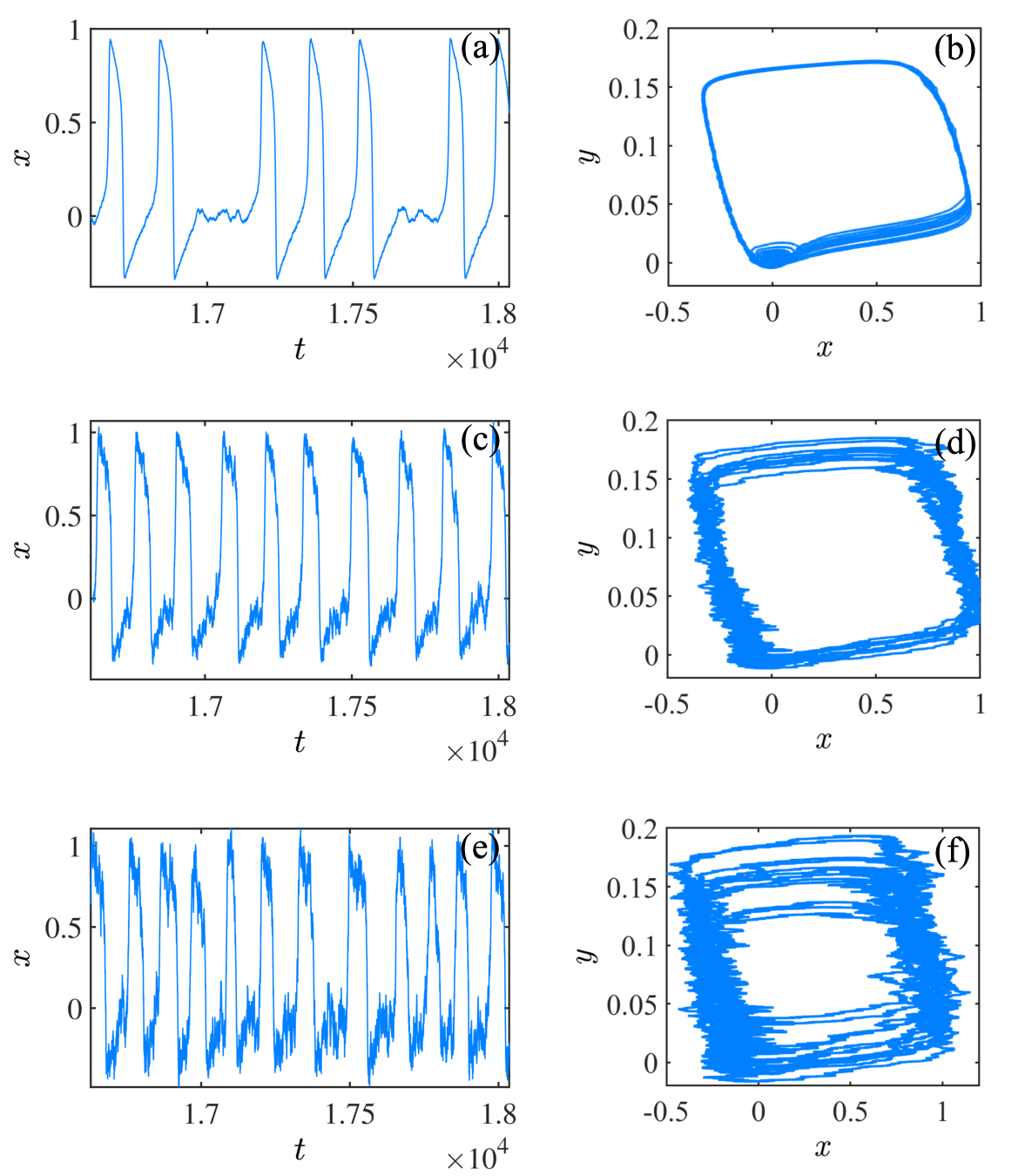}
		\caption{\label{cvt}Time series and phase portraits illustrating system dynamics at different noise amplitudes: For  $D = 0.001$  [(a)-(b)], infrequent spikes are observed in the time series, with  $CV = 1$ , indicating less regular oscillations. At  $D = 0.1$  [(c)-(d)], the time series shows frequent spiking activity, while the phase portrait exhibits a stochastic limit cycle, corresponding to  $CV = 0.15$. For $D = 0.3$ [(e)-(f)], aperiodic and infrequent spikes are observed in both the time series and phase portrait, accompanied by an increase in CV value }
	\end{figure}
	
	To quantify the system’s response to noise and spike frequency, we employ the well-known coefficient of variation (CV) statistic for interspike intervals (ISI) \cite{pikovsky1997coherence,deville2005sisr}. The CV is calculated as:
	\begin{equation}
		CV = \frac{\sqrt{\langle ISI^2 \rangle - \langle ISI \rangle^2}}{\langle ISI \rangle}.
	\end{equation}
	
This statistic serves as a critical tool for distinguishing the various dynamical behaviors exhibited by the system. Primarily, it is employed to detect and characterize the periodic responses of the system under varying noise intensities. A higher $CV$ indicates increased irregularity in spike timing, with $CV \geq 1$ suggesting a Poisson-like, random process. Interspike intervals were measured on the variable $x$ using a spike detection threshold of 0.8. Figure \ref{cv} illustrates how $CV$ varies with noise intensity $D$ over the range $D \in (0.0001, 0.5)$ on a logarithmic scale, identifying distinct regions (R1–R5) that represent different spiking behaviors.
	
	In the absence of noise ($D < 0.0002$, region R1), no spikes are detected, and consequently, no $CV$ values are recorded. As noise intensity reaches $D = 0.0002$ (region R2), rare events begin to occur, resulting in a $CV$ value greater than 1, which signifies incoherent and independent spiking events. As $D$ increases further, the $CV$ decreases, reaching $CV = 1$ at $D = 0.001$, signifying random spiking behavior. This transition is visible in the time series in Figs. \ref{cvt}(a-b), where the infrequent spikes are consistent with the observed $CV$ values, with the bursting behavior discussed previously falling within region R3.
	
	With further increases in $D$, the $CV$ continues to decrease, reaching a minimum value of 0.15 at $D = 0.1$ (region R4), indicating the onset of coherent spike production and resonance. This behavior is depicted in Fig. \ref{cv}, and Figs. \ref{cvt}(c-d) illustrate the frequent oscillations observed at $D = 0.1$, where a stochastic limit cycle appears in the phase portrait. This coherence, attributed to SISR, arises when the intrinsic frequency of the system aligns with the noise, producing resonant oscillations. Notably, these oscillations are entirely noise-dependent. In our study, system parameters are specifically set to maintain bifurcation thresholds distant from fixed points, ensuring the absence of limit cycles without noise. Consequently, these observed oscillations are purely noise-driven and categorized as SISR.
	
	Beyond $D = 0.1$, $CV$ begins to increase gradually, leading to incoherent spiking once more. At this stage, the higher noise intensity disrupts the resonance, causing the system to oscillate randomly, as shown in region R5. This behavior is evident in Figs. \ref{cvt}(e-f) for $D = 0.3$. Thus, the system exhibits rare, high-amplitude EE at low noise intensities, transitions through random spiking oscillations, achieves a resonance peak at $D = 0.1$, and eventually reverts to incoherent oscillations as noise intensity increases further.This progression underscores the phenomenon that resonance occurs at higher noise intensities, providing further support for the concept that noise alone can induce periodic oscillations, as demonstrated in previous studies \cite{rappel1994stochastic}.
	
	\section{Mechanism of extreme events }
	The system under investigation exhibits spiking behavior in response to noise, where at lower noise intensities, these spikes correspond to extreme and rare events, while at higher noise intensities, they become more regular bursting. To understand the mechanism underlying the emergence of EE due to noise, we employ the large fluctuations approach. This method, as outlined in \cite{khovanov2013noise}, has been applied to a variant of the FHN model to analyze how trajectories escape from the vicinity of stable points. As noted earlier, for our analysis, the system must remain in a non-self-oscillating regime, enabling us to isolate the role of noise in generating EE. We follow a similar algorithm to the approach described in \cite{khovanov2013noise, beri2005solution} to investigate how this process leads to the formation of EE.
	\begin{figure}[h!]
		\centering
		\includegraphics[width=1.0\linewidth]{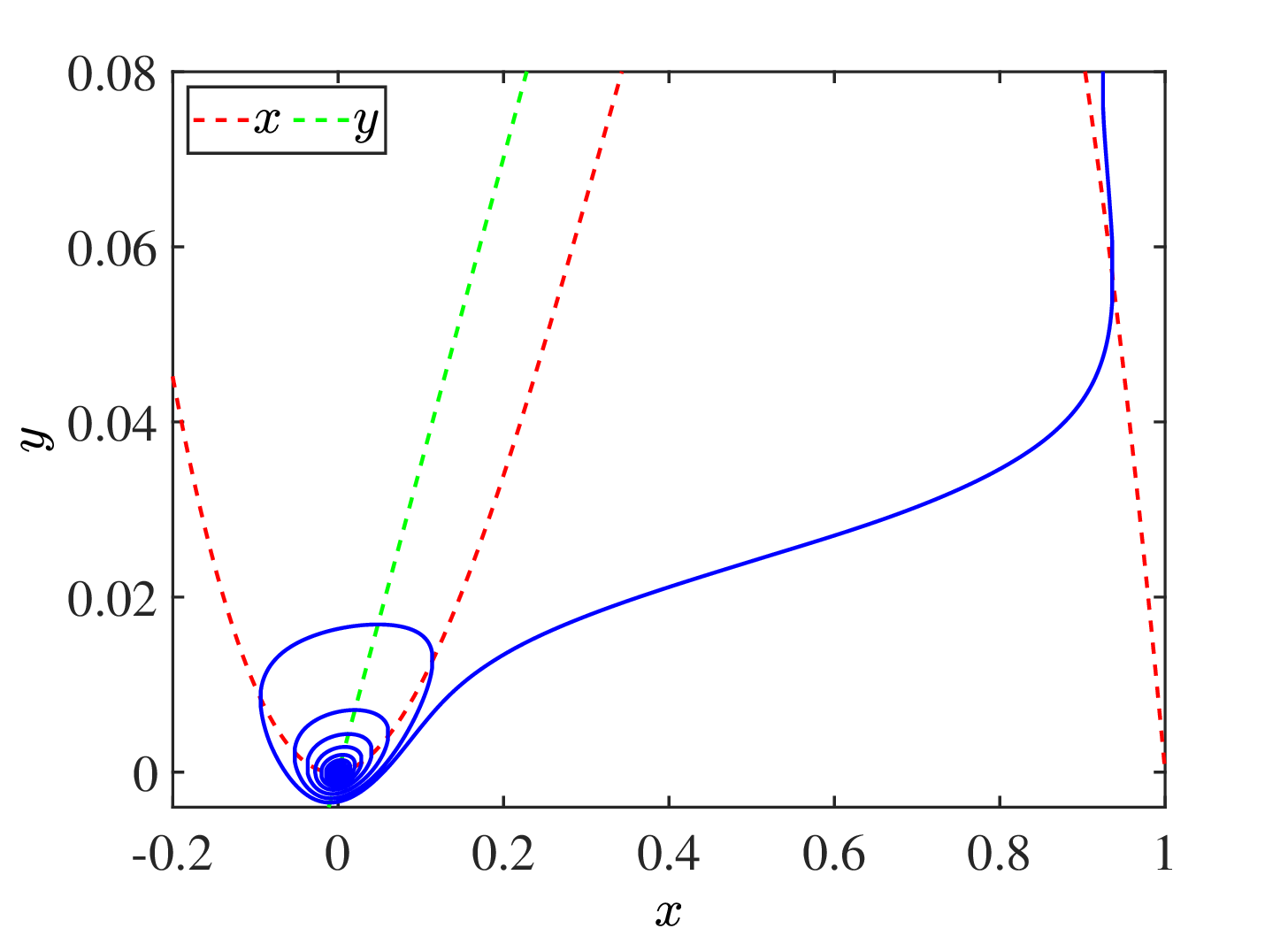}
		\caption{\label{enull} Phase space of Eq. (\ref{eq:e7}), illustrating the presence of a fixed point identified by the intersection of nullclines ( $x$-axis: red dashed line;  $y$-axis: green dashed line). The trajectories evolve from the vicinity of the fixed point, becoming unstable and following minimal energy pathways during their excursions.}
	\end{figure}
	\par
	
	We begin by considering our system, represented by Eq. (\ref{eq:e2}), which is influenced by noise. The corresponding Fokker-Planck equation, describing the probability density $P(x, y, t)$ for the system, is given by:
	\begin{dmath}
		\label{eq:e3}
		\frac{\partial P(x,y,t)}{\partial t} =  \frac{\partial}{\partial x} \left[-( x(a-x)(x-1)-y) P \right] +\frac{\partial}{\partial y} \left[ -(bx-cy) P \right]+ D \frac{\partial^2 P}{\partial x^2}  .
	\end{dmath}
	The solution to this equation is expressed using a WKB-like approximation as:
	\begin{eqnarray}
		\label{eq:e4}
		P(x,y) = Z(x,y) \exp \left[ \frac{-S(x,y)}{D} \right].
	\end{eqnarray} 
	Here, $Z(x, y)$ is a prefactor, and $S(x, y)$ represents the quasipotential, or the action, for the system described by Eq. (\ref{eq:e2}). Inserting this expression into the Fokker-Planck equation and following the procedure outlined in \cite{khovanov2013noise}, we arrive at the leading order approximation:
	\begin{eqnarray}
		\label{eq:e5}
		\dot{x}= \frac{\partial H}{\partial p_x},    \dot{y}= \frac{\partial H}{\partial p_y}, \dot{p_x}= -\frac{\partial H}{\partial x}, \dot{p_y}= -\frac{\partial H}{\partial y},
	\end{eqnarray} 
	where the Hamiltonian $H$ takes the Wentzel-Freidlin form:
	\begin{eqnarray}
		\label{eq:e6}
		H = p_x(x(a-x)(x-1)-y)+p_y(bx-cy)+\frac{p^2_x}{2}.
	\end{eqnarray} 
	Here, $p_x$ and $p_y$ represent the conjugate momenta. Using this Hamiltonian, we derive the Hamilton’s equations of motion as:
	\begin{eqnarray}
		\label{eq:e7}
		\dot{x}&=& x(a-x)(x-1)-y+p_x, \nonumber \\
		\dot{y} &=& bx-cy, \nonumber \\
		\dot{p_x} &=& -(2ax-a-3x^2+2x)p_x + p_y b,\\
		\dot{p_y}&=&p_x +c p_y. \nonumber 
	\end{eqnarray} 
	These equations identify the most probable trajectories, which connect the system’s initial and final states under the influence of noise, with minimal energy or action ($S$). The action is given by:
	\begin{eqnarray}
		\label{eq:e8}
		S = min \int_{t_i}^{t_f} \xi_x^2(t) \, dt, 
	\end{eqnarray}
	
	\begin{eqnarray}
		\label{eq:e9}
		S = min \hspace{0.1cm}\frac{1}{2}\int_{t_i}^{t_f} p_x^2 (t) \, dt. 
	\end{eqnarray}
	
	Here, the action $S$ quantifies the energy required for a trajectory to escape the stable region and is minimized to determine the most probable and minimal energy pathway. Since we treat noise as analogous to momentum in the Hamiltonian, we substitute $\xi^2(t) = \frac{p_x^2}{2}$. To find the trajectory that minimizes the action, we solve a boundary value problem with the following boundary conditions: $[x_i(t_i), y_i(t_i), p_x(t_i) = 0, p_y(t_i) = 0]$ at the initial time $t_i, and [x_f(t_f), y_f(t_f), p_x(t_f) = 0, p_y(t_f) = 0]$ at the final time $t_f$. The final time $t_f$ is unspecified, and we use a shooting method to determine the escape path \cite{beri2005solution}.
	\begin{figure*}[h!]
		\centering
		\includegraphics[width=1.0\linewidth]{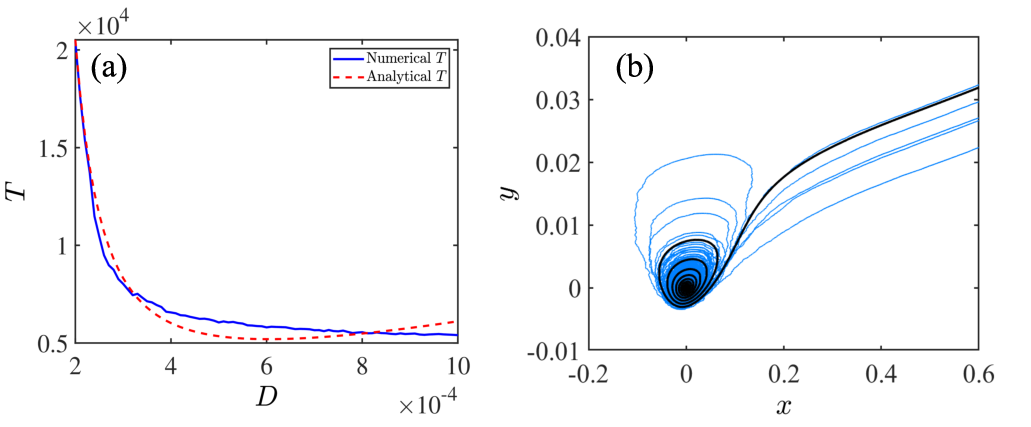}
		\caption{\label{er}(a) Numerically estimated escape rate $T$ (solid blue line) based on 200 realizations for each noise intensity, compared with the analytical prediction (dashed red line) derived from the estimated  $S$ . The analytical and numerical results exhibit strong agreement at lower noise intensities, where EE emerge. (b) Evolution of Eq. (\ref{eq:e2}) for multiple runs from identical initial conditions at the stable fixed point ($x, y$) = ($0, 0$) for  $D = 0.0002$, where EE emerge.Each trajectory departs from the fixed point, undergoing excursions in phase space. The thick black line represents the evolution of Eq. (\ref{eq:e7}), illustrating the minimal energy pathway needed to cross the stable fixed point.}
	\end{figure*}
	
	In \cite{khovanov2013noise}, the authors describe the FitzHugh-Nagumo (FHN) system, which features a ghost separatrix. This structure, lacking a saddle point, implies that the system cannot inherently produce spikes without external noise. The ghost separatrix is a remnant of transient behavior after a bifurcation event, where the saddle and node annihilate each other and vanish. What remains is a “ghosted” boundary in phase space that transiently influences nearby trajectories. In contrast, our system features a real separatrix, defined by a saddle point and a stable fixed point. This separatrix plays a persistent role, permanently dividing the phase space into regions with distinct long-term dynamics. Depending on the system parameters, significant excursions in phase space can occur under perturbation. For instance, if the initial condition is set near the stable fixed point, the system remains in the stable regime in the absence of perturbations (i.e., when the momenta are zero), always returning to the stable point and never reaching the saddle. However, introducing a small perturbation in $p_x$ or $p_y$ enables the system to escape the stable regime. This escape allows the saddle to direct the trajectory into a spike. In the Hamiltonian framework, the momenta ($p_x$, $p_y$) play a role analogous to noise in the original FHN system.
	
	In our system, the saddle point is located far from the stable fixed point. This separation means that the system spends significant time oscillating around the stable point before gaining enough energy to meet the saddle and produce an EE. Figure \ref{enull} illustrates the phase space trajectories (solid lines) of Eq. (\ref{eq:e7}), with the dashed lines representing the system’s nullclines. Even when the ICs are chosen in the stable regime, a minimal perturbation in the momentum causes the system to oscillate around the fixed point for a prolonged period before crossing the boundary, leading to EE emergence.
	
	If the parameters are chosen such that the saddle moves closer to the stable point, the system requires less energy to cross the saddle, resulting in more frequent spike production. Given the current parameter set, where the saddle is far from the stable point, we can calculate the action $S$ as outlined in Eq. (\ref{eq:e9}) to quantify the emergence of EE.
	
	\subsection{Escape rate of trajectories}
	Having determined the action $S$, we now proceed to estimate the escape rate numerically. This serves as a measure of the system’s stability and provides a means to validate the action $S$, which corresponds to the optimal paths for overcoming the stable fixed point. To achieve this, we simulate Eq. (\ref{eq:e2}) and calculate the time interval $\tau$ for the trajectory to initiate an excursion or spike, starting from a specified initial condition. This process is repeated across multiple realizations to obtain the average escape rate, defined as $T = \frac{1}{N} \sum_{i=1}^{N} \tau_i$, where $N = 200$. The average escape rate $T$ can also be expressed in terms of the action $S$ according to the relationship $T \propto D^{3/2} \exp\left(\frac{S}{D}\right).$
	
	To account for the dependence of this proportional relationship on $D$, we employ a power-law fitting approach to scale the analytical results. As shown in Fig. \ref{er} (a), the analytical escape rate  $T$  closely aligns with the numerical escape rate calculated from Eq. (\ref{eq:e2}) for small values of  $D$ . The figure illustrates that the escape rate is relatively high due to the rarity of EE. As the noise intensity increases, $T$ decreases. This provides clear evidence that higher noise levels facilitate faster movement of trajectories away from the fixed point. This observation aligns with the expectation that spikes tend to occur more frequently under conditions of increased noise intensity.
	
	The escape paths of the trajectories for  $D = 0.0002$  are depicted in Fig.\ref{er} (b), where the blue lines represent the solutions of Eq. (\ref{eq:e2}) across different runs. Each trajectory follows a unique path and requires a considerable amount of time to escape the vicinity of the fixed point. The thick black line represents the solution of the Hamiltonian equations, which similarly requires a significant duration to achieve an excursion from the stable point. Thus, we conclude that for EE to manifest due to low-intensity noise, the escape rate is elevated, consistent with minimal perturbations in momentum within the Hamiltonian framework. Additionally, we have successfully computed the minimal action path that characterizes the trajectory of EE.
	
	\section{Conclusion}
	Our research aims to advance the understanding of the FHN model and investigate the relatively under-researched phenomenon of noise-induced EE. By focusing on both deterministic and stochastic aspects of the system, we isolated key mechanisms to clarify the differences introduced by noise. Given the stochastic nature of the system, we employed probabilistic measures to quantify these noise-induced changes. This ultimately led to the occurrence of EE through a P-bifurcation analysis. Importantly, our findings highlight that EE emerge solely as a response to noise, and the pathways leading to these events have been clarified via bifurcation analysis. In addition to EE, our study identified other noise-induced phenomena, such as noise-induced bursting at specific noise intensities, which we analyzed using ISI statistics. To provide a broader context for the system’s behavior, we incorporated the CV as a measure, linking our results to previously documented dynamical phenomena in the FHN model.
	
	A noteworthy observation was the occurrence of frequent spikes in the system, which can be attributed to self-induced stochastic resonance. Moreover, we applied large fluctuation theory to examine the emergence of EE in noise-driven systems, calculating escape rates of system trajectories using both analytical and numerical approaches. This combined methodology provides valuable insights into the mechanisms driving EE in stochastic systems.
	
	In conclusion, our investigation deepens the understanding of EE in stochastic systems, particularly within the context of neuronal models. The insights gained from this study lay the foundation for further research and potential applications in neuronal science and other fields where noise-induced dynamics play a significant role.
	
	\section*{Acknowledgement}
	The research contributions of S.H., R.S., and V.K.C. are part of a project funded by the SERB-CRG (Grant No. CRG/2022/004784). The authors gratefully acknowledge the Department of Science and Technology (DST), New Delhi, for providing computational facilities through the DST-FIST program under project number SR/FST/PS-1/2020/135 , awarded to the Department of Physics.\\
	
	\noindent {\bf Author Contributions} All the authors contributed equally to the preparation of this manuscript.\\
	
	\noindent {\bf Data Availability Statement} The authors confirm that the data supporting the findings of this study are available within the article.\\

	\bibliographystyle{elsarticle-num} 
	\bibliography{cas-refs1}

\begin{thebibliography}{10}
\expandafter\ifx\csname url\endcsname\relax
  \def\url#1{\texttt{#1}}\fi
\expandafter\ifx\csname urlprefix\endcsname\relax\def\urlprefix{URL }\fi
\expandafter\ifx\csname href\endcsname\relax
  \def\href#1#2{#2} \def\path#1{#1}\fi

\bibitem{matsumoto1983noise}
K.~Matsumoto, I.~Tsuda, Noise-induced order, Journal of Statistical Physics 31
  (1983) 87--106.

\bibitem{Gassman1997noise}
F.~Gassmann, Noise-induced chaos-order transitions, Physical Review E 55~(3)
  (1997) 2215.

\bibitem{werner1984noise}
W.~Horsthemke, Noise induced transitions, in: Non-Equilibrium Dynamics in
  Chemical Systems: Proceedings of the International Symposium, Bordeaux,
  France, September 3--7, 1984, Springer, 1984, pp. 150--160.

\bibitem{gao1999noise_chaos}
J.~Gao, S.~Hwang, J.~Liu, When can noise induce chaos?, Physical review letters
  82~(6) (1999) 1132.

\bibitem{duan2015introduction}
J.~Duan, An introduction to stochastic dynamics, Vol.~51, Cambridge University
  Press, 2015.

\bibitem{tsimring2014noise}
L.~S. Tsimring, Noise in biology, Reports on Progress in Physics 77~(2) (2014)
  026601.

\bibitem{fishmm1990electrical}
H.~FiSHMM, H.~Leuchtag, Electrical noise in physics and biology, in: Current
  topics in membranes and transport, Vol.~37, Elsevier, 1990, pp. 3--35.

\bibitem{faisal2008noise}
A.~A. Faisal, L.~P. Selen, D.~M. Wolpert, Noise in the nervous system, Nature
  reviews neuroscience 9~(4) (2008) 292--303.

\bibitem{osorio2010}
D.~S. J.~M. Ivan~Osorio, Mark G.~Frei, Y.-C. Lai, Epileptic seizures: Quakes of
  the brain?, Physical Review E 82 (2010) 021919.

\bibitem{lehnertz2006}
K.~Lehnertz, Epilepsy: Extreme events in the human brain, in: Extreme Events in
  Nature and Society. The Frontiers Collection, Springer, Berlin, Heidelberg,
  2006, Springer, 2006, pp. 123--143.

\bibitem{lebedev2006}
N.~M. A.~L. Lebedev, M~A., Brain-machine interfaces: Past, present and future,
  Trends in Neurosciences 29~(9) (2006) 536--546.

\bibitem{bottou2010}
L.~Bottou, Large-scale machine learning with stochastic gradient descent,
  Proceedings of COMPSTAT'2010 (2010) 177--186.

\bibitem{hodgkin1952}
A.~F.~H. A~L~Hodgkin, A quantitative description of membrane current and its
  application to conduction and excitation in nerve, The Journal of Physiology
  117~(4) (1952) 500--544.

\bibitem{fitzhugh1961}
R.~FitzHugh, Impulses and physiological states in theoretical models of nerve
  membrane, Biophysical journal 1~(6) (1961) 445--466.

\bibitem{sgro2015intracellular}
A.~E. Sgro, D.~J. Schwab, J.~Noorbakhsh, T.~Mestler, P.~Mehta, T.~Gregor, From
  intracellular signaling to population oscillations: bridging size-and
  time-scales in collective behavior, Molecular systems biology 11~(1) (2015)
  779.

\bibitem{cebrian2024six}
D.~Cebr{\'\i}an-Lacasa, P.~Parra-Rivas, D.~Ruiz-Reyn{\'e}s, L.~Gelens, Six
  decades of the fitzhugh-nagumo model: A guide through its spatio-temporal
  dynamics and influence across disciplines, arXiv preprint arXiv:2404.11403
  (2024).

\bibitem{makarov2001spiking}
V.~A. Makarov, V.~I. Nekorkin, M.~G. Velarde, Spiking behavior in a
  noise-driven system combining oscillatory and excitatory properties, Physical
  Review Letters 86~(15) (2001) 3431.

\bibitem{pikovsky1997coherence}
A.~S. Pikovsky, J.~Kurths, Coherence resonance in a noise-driven excitable
  system, Physical Review Letters 78~(5) (1997) 775.

\bibitem{longtin1997sr}
A.~Longtin, Autonomous stochastic resonance in bursting neurons, Physical
  Review E 55~(1) (1997) 868.

\bibitem{muratov2008mmo}
C.~B. Muratov, E.~Vanden-Eijnden, Noise-induced mixed-mode oscillations in a
  relaxation oscillator near the onset of a limit cycle, Chaos: An
  Interdisciplinary Journal of Nonlinear Science 18~(1) (2008).

\bibitem{manchein2022noise}
C.~Manchein, L.~Santana, R.~M. da~Silva, M.~W. Beims, Noise-induced
  stabilization of the fitzhugh--nagumo neuron dynamics: Multistability and
  transient chaos, Chaos: An Interdisciplinary Journal of Nonlinear Science
  32~(8) (2022).

\bibitem{bashkirtseva2014noise}
I.~Bashkirtseva, L.~Ryashko, E.~Slepukhina, Noise-induced oscillating
  bistability and transition to chaos in fitzhugh--nagumo model, Fluctuation
  and noise letters 13~(01) (2014) 1450004.

\bibitem{muratov2005sisr}
C.~B. Muratov, E.~Vanden-Eijnden, E.~Weinan, Self-induced stochastic resonance
  in excitable systems, Physica D: Nonlinear Phenomena 210~(3-4) (2005)
  227--240.

\bibitem{deville2005sisr}
R.~L. DeVille, E.~Vanden-Eijnden, C.~B. Muratov, Two distinct mechanisms of
  coherence in randomly perturbed dynamical systems, Physical Review E 72~(3)
  (2005) 031105.

\bibitem{wu2024resonance}
Y.~Wu, Z.~Sun, N.~Zhao, Resonance dynamics in multilayer neural networks
  subjected to electromagnetic induction, Communications in Nonlinear Science
  and Numerical Simulation (2024) 108575.

\bibitem{xu2020dynamics}
Y.~Xu, Y.~Guo, G.~Ren, J.~Ma, Dynamics and stochastic resonance in a
  thermosensitive neuron, Applied Mathematics and Computation 385 (2020)
  125427.

\bibitem{helbing2001social}
D.~Helbing, Traffic and related self-driven many-particle systems, Reviews of
  Modern Physics 73~(4)  1067.

\bibitem{chabchoub2011ocean}
A.~Chabchoub, N.~Hoffmann, N.~Akhmediev, Rogue wave observation in a water wave
  tank, Physical Review Letters 106~(20) (2011) 204502.

\bibitem{bonatto2011optics}
C.~Bonatto, M.~Feyereisen, S.~Barland, M.~Giudici, C.~Masoller, J.~R.~R. Leite,
  J.~R. Tredicce, Deterministic optical rogue waves, Physical review letters
  107~(5) (2011) 053901.

\bibitem{sornette2006geo}
D.~Sornette, Critical phenomena in natural sciences: chaos, fractals,
  selforganization and disorder: concepts and tools, Springer Science \&
  Business Media, 2006.

\bibitem{bunde2012eco}
A.~Bunde, J.~Kropp, H.-J. Schellnhuber, The science of disasters: climate
  disruptions, heart attacks, and market crashes, Springer Science \& Business
  Media, 2012.

\bibitem{dobson2007power}
I.~Dobson, B.~A. Carreras, V.~E. Lynch, D.~E. Newman, Complex systems analysis
  of series of blackouts: Cascading failure, critical points, and
  self-organization, Chaos: An Interdisciplinary Journal of Nonlinear Science
  17~(2) (2007).

\bibitem{Gupta2020SingleNeuron}
P.~Gupta, N.~Balasubramaniam, H.~Y. Chang, F.~G. Tseng, T.~S. Santra, A
  single-neuron: Current trends and future prospects, Cells 9~(6) (2020) 1528.

\bibitem{hariharan2024noise}
S.~Hariharan, R.~Suresh, V.~Chandrasekar, Noise-induced extreme events in
  integer and fractional-order memristive hindmarsh--rose neuron models: a
  comprehensive study, The European Physical Journal Plus 139~(3) (2024) 292.

\bibitem{vijay2023ses}
S.~D. Vijay, K.~Thamilmaran, A.~I. Ahamed, Superextreme spiking oscillations
  and multistability in a memristor-based hindmarsh--rose neuron model,
  Nonlinear Dynamics 111~(1) (2023) 789--799.

\bibitem{olenin2023hree}
S.~M. Olenin, T.~A. Levanova, Extreme events in small ensemble of bursting
  neurons with chemical and electrical couplings, in: 2023 International Joint
  Conference on Neural Networks (IJCNN), IEEE, 2023, pp. 1--6.

\bibitem{karnatak2014fnee}
R.~Karnatak, G.~Ansmann, U.~Feudel, K.~Lehnertz, Route to extreme events in
  excitable systems, Physical Review E 90~(2) (2014) 022917.

\bibitem{saha2017fndelay}
A.~Saha, U.~Feudel, Extreme events in fitzhugh-nagumo oscillators coupled with
  two time delays, Physical Review E 95~(6) (2017) 062219.

\bibitem{chowdhury2022review}
S.~N. Chowdhury, A.~Ray, S.~K. Dana, D.~Ghosh, Extreme events in dynamical
  systems and random walkers: A review, Physics Reports 966 (2022) 1--52.

\bibitem{MATLABfindpeaks}
I.~The~MathWorks,
  \href{https://in.mathworks.com/help/signal/ref/findpeaks.html}{Find Local
  Maxima - MATLAB findpeaks}, accessed: 2024-10-20 (2023).
\newline\urlprefix\url{https://in.mathworks.com/help/signal/ref/findpeaks.html}

\bibitem{arnold1995rds}
L.~Arnold, C.~K. Jones, K.~Mischaikow, G.~Raugel, L.~Arnold, Random dynamical
  systems, Springer, 1995.

\bibitem{bashkirtseva2018methods}
I.~Bashkirtseva, L.~Ryashko, E.~Slepukhina, Methods of stochastic analysis of
  complex regimes in the 3d hindmarsh--rose neuron model, Fluctuation and Noise
  Letters 17~(01) (2018) 1850008.

\bibitem{channell2009burst}
P.~Channell, I.~Fuwape, A.~B. Neiman, A.~L. Shilnikov, Variability of bursting
  patterns in a neuron model in the presence of noise, Journal of Computational
  Neuroscience 27 (2009) 527--542.

\bibitem{ryashko2020burst}
L.~Ryashko, E.~Slepukhina, Noise-induced toroidal excitability in neuron model,
  Communications in Nonlinear Science and Numerical Simulation 82 (2020)
  105071.

\bibitem{lopez2023controlling}
J.~L{\'o}pez, M.~Coccolo, R.~Cape{\'a}ns, M.~A. Sanju{\'a}n, Controlling the
  bursting size in the two-dimensional rulkov model, Communications in
  Nonlinear Science and Numerical Simulation 120 (2023) 107184.

\bibitem{baltanas1998burst}
J.~Baltanas, J.~Casado, Bursting behaviour of the fitzhugh-nagumo neuron model
  subject to quasi-monochromatic noise, Physica D: Nonlinear Phenomena
  122~(1-4) (1998) 231--240.

\bibitem{longtin1995synchronization}
A.~Longtin, Synchronization of the stochastic fitzhugh-nagumo equations to
  periodic forcing, Il Nuovo Cimento D 17 (1995) 835--846.

\bibitem{rappel1994stochastic}
W.-J. Rappel, S.~H. Strogatz, Stochastic resonance in an autonomous system with
  a nonuniform limit cycle, Physical Review E 50~(4) (1994) 3249.

\bibitem{khovanov2013noise}
I.~Khovanov, A.~Polovinkin, P.~McClintock, Noise-induced escape in an excitable
  system, Physical Review E 87~(3) (2013) 032116.

\bibitem{beri2005solution}
S.~Beri, R.~Mannella, D.~Luchinsky, A.~Silchenko, P.~McClintock, Solution of
  the boundary value problem for optimal escape in continuous stochastic
  systems and maps, Physical Review E 72~(3) (2005) 036131.

\end{thebibliography}
	
	
		
		
		
	
\end{document}